\documentclass[12pt]{article}
\usepackage{amssymb,amsfonts}
\newcommand{\be}{\begin{eqnarray}}
\newcommand{\ee}{\end{eqnarray}}
\newcommand{\A}{{\cal A}}
\newcommand{\R}{{\cal R}}
\newcommand{\pa}{\partial}
\renewcommand{\d}{{\rm d}}
\newcommand{\td}{\tilde{\d}}
\newcommand{\tA}{\tilde{A}}
\newcommand{\D}{{\rm D}}
\newcommand{\dl}{{\delta}}
\newcommand{\pat}[1]{{\pa #1\over\pa\vartheta}}

\title{\bf Moyal Deformation, Seiberg-Witten-Map, and Integrable Models}
\date{  }
\author{A. Dimakis \\ Department of Mathematics, University of the Aegean \\
        GR-83200 Karlovasi, Samos, Greece \\ dimakis@aegean.gr
        \\[2ex]
        F. M\"uller-Hoissen \\ Max-Planck-Institut f\"ur Str\"omungsforschung \\
        Bunsenstrasse 10, D-37073 G\"ottingen, Germany \\
        fmuelle@gwdg.de }

\begin{document}
\renewcommand{\theequation} {\arabic{section}.\arabic{equation}}

\maketitle

\begin{abstract}
A covariant formalism for Moyal deformations of gauge theory and differential 
equations which determine Seiberg-Witten maps is presented. 
Replacing the ordinary product of functions by the noncommutative
Moyal product, noncommutative versions of integrable models can
be constructed. We explore how a Seiberg-Witten map acts in such a
framework. As a specific example, we consider a noncommutative extension
of the principal chiral model.
\end{abstract}

\section{Introduction}
\setcounter{equation}{0}
 Field theory on noncommutative spaces has more and more attracted the
attention of researchers during the last years. A major impulse
came from the discovery that a noncommutative gauge field theory
arises in a certain limit of string theory (see \cite{Seib+Witt99} 
and the references cited there). 
In \cite{Seib+Witt99} a (perturbative) equivalence between ordinary
and noncommutative gauge fields was established via a change of variables
to which the name {\em Seiberg-Witten map} was assigned in subsequent 
publications \cite{Asak+Kish99,sw}. More generally, models on noncommutative space-times 
obtained by replacing the ordinary product of functions by the 
noncommutative Moyal product \cite{dq} were explored in several recent 
publications. In particular, Moyal deformations of integrable models 
were constructed via deformation of an associated bicomplex 
\cite{DMH00a,DMH00b,DMH00c}. 
\vskip.1cm

Section 2 collects some notes on deformations of products and, in particular, 
recalls the definition of the Moyal $\ast$-product. Section 3 deals with a 
corresponding deformation of gauge theory, develops a covariant differentiation 
formalism with respect to deformation parameters, generalizes the Seiberg-Witten 
map \cite{Seib+Witt99} from infinitesimal to finite gauge transformations, and 
shows that this map describes a parallel transport along a curve in the 
deformation parameter space. Noncommutativity of covariant derivatives with 
respect to different deformation parameters is associated with a notion of 
curvature in section 4. In section 5 we show how the Seiberg-Witten map can 
be used to generate from solutions of a classical integrable model solutions 
of the corresponding deformed noncommutative model. As a specific example, 
we consider a noncommutative version of the principal chiral model. Section 6  
contains some concluding remarks.

\section{Deformations of products}
\setcounter{equation}{0}
We take the opportunity to make a few remarks about deformations of products 
and present some useful formulas and instructive examples. Much of it is not 
really needed in the following sections, however.
\vskip.1cm
  
Let $\A$ be an associative unital algebra over a commutative ring $\R$ and
let $m : \A \otimes \A \to \A$ denote\footnote{Here and in what follows 
we simply write $\otimes$ instead of $\otimes_\R$.}
the multiplication in $\A$. Given a map $R : \A \otimes \A \to A \otimes \A$, 
we define a deformed multiplication in $\A$ by
\be
   f \ast g = \hat{m} (f \otimes g)   \, , \qquad
   \hat{m} = m \, R   \; .
\ee
In general, this is not an associative deformation. If we assume that
$R$ satisfies\footnote{These are duals of the relations which define
quasitriangular Hopf algebras.}
\be
   R \, m_{12} = m_{12} \, R_{23} \, R_{13} \, , \qquad
   R \, m_{23} = m_{23} \, R_{12} \, R_{13} \, ,
                    \label{tri}
\ee
then $\ast$ is associative if $R$ also satisfies the Yang-Baxter equation                    
\be
    R_{12} \, R_{13} \, R_{23} = R_{23} \, R_{13} \, R_{12}  \; .
                   \label{y-b}
\ee
Indeed, it is easy to check associativity,\footnote{Actually, (\ref{assoc}) 
shows that the weaker conditions $m \, R \, m_{12} = m \, m_{12} \, R_{23} \, R_{13}$, 
$m \, R \, m_{23} = m \, m_{23} \, R_{12} \, R_{13}$ and 
$m \, m_{12} R_{12} R_{13} R_{23} = m \, m_{12} R_{23} R_{13} R_{12}$ 
are sufficient to ensure associativity.}     
\be
 \hat{m} \, \hat{m}_{12} &=& m \, R \, m_{12} R_{12} 
                          = m \, m_{12} R_{23} R_{13} R_{12}    
                          = m \, m_{12} R_{12} R_{13} R_{23} \nonumber \\
                         &=& m \, m_{23} R_{12} R_{13} R_{23} 
                          = m \, R \, m_{23} R_{23}           
                          = \hat{m} \, \hat{m}_{23}
                            \label{assoc}
\ee
using the associativity condition $m \, m_{12} = m \, m_{23}$ for $m$.
\vskip.1cm

Two $\ast$-products $\ast,\,\ast'$ will be considered to be
equivalent, if there is an invertible linear map $S : \A \to \A$ such that  
\be
    S \, \hat{m} = \hat{m}' \, S \otimes S  
\ee
which is equivalent to
\be
    S(f \ast g) = (S f) \ast' (S g) \; . \label{eqst}
\ee
If $S$ is also an $(\A,m)$-automorphism, i.e., $S \, m = m \, (S \otimes S)$, 
the above condition reads $m \, (S \otimes S) \, R = S \, m \, R = m \, R 
\, (S \otimes S)$ which is satisfied if $(S \otimes S) R = R' \, (S \otimes S)$.
\vskip.1cm

As an example, let us define
\be
   P = \theta^{ij} \, \pa_i \otimes \pa_j \, , \qquad
   R = e^{P/2} = I + {1 \over 2} \, P + \cdots
\ee
where $\pa_i : \A \to \A$ are commuting derivations of $\A$,
$\theta^{ij} \in \R$, and $I : \A \to \A$ is the identity operator.
Using the derivation property
\be
   \pa_i \, m = m \, (\pa_i \otimes I + I \otimes \pa_i) 
\ee
we get
\be
   P \, m_{12} 
 = \theta^{ij} \, \pa_i \, m_{12} \otimes \pa_j 
 = \theta^{ij} \,  m_{12} \, (\pa_i \otimes I + I \otimes \pa_i) \otimes \pa_j 
 = m_{12} \, (P_{13} + P_{23}) 
\ee
and in the same way
\be
   P \, m_{23} = m_{23} \, (P_{12} + P_{13})  \; .
\ee
It follows that (\ref{tri}) and (\ref{y-b}) are satisfied.
Note that $\pa_i$ are also derivations of $\A$ with respect to
the product $\hat{m}$. $\A$ need not be commutative.
\vskip.1cm

If $\A$ is the algebra of smooth functions on $\mathbb{R}^{2n}$ and
$\theta^{ij} = i \, h\, \epsilon^{ij}$ with real $\epsilon^{ij}$ is antisymmetric
and nondegenerate, then $\hat{m}$ is the well-known Moyal product \cite{dq}.
\vskip.1cm

If $\A$ is the Heisenberg algebra with $[x^i,x^j] = i \, h \, \epsilon^{ij} I$, 
then ${\rm ad}(x^i) y = [x^i , y]$ defines commuting derivations. Let
\be
   P = {1 \over i \, h} \, \epsilon_{ij} \, {\rm ad}(x^i) \otimes {\rm ad}(x^j)
\ee
where $\epsilon_{ij}$ is antisymmetric and satisfies
$\epsilon_{ik} \, \epsilon^{jk} = \delta^j_i$. Then one easily verifies that
$x^i \ast x^j = x^j \ast x^i$ and $\hat{m}$ is commutative.
\vskip.1cm

As a further example, consider the algebra $\A=M(3,\mathbb{C})$ of
$3\times 3$-matrices with complex coefficients. Let $E_{ij}$ denote 
the matrix with entry $1$ in the $i$th row and $j$th column and 
otherwise $0$. Setting $H_1 = E_{11}-E_{22}$ and $H_2 = E_{22}-E_{33}$, 
the derivations ${\rm ad}(H_i)$, $i=1,2$, of $\A$ commute with each other. 
With $P = \vartheta \, [{\rm ad}(H_1) \otimes {\rm ad}(H_2)-{\rm ad}(H_2)
\otimes {\rm ad}(H_1)]$ we get the following associative deformation
of the ordinary matrix multiplication,
\be
  \begin{array}{ccccc}
  E_{12} \ast E_{23} = q E_{13}, & \qquad & E_{23} \ast E_{31} = q E_{21}
  & \qquad & E_{31} \ast E_{12} = q E_{32} \\
  E_{13} \ast E_{32} = q^{-1} E_{12} & & E_{21} \ast E_{13} = q^{-1} E_{23}
  & & E_{32} \ast E_{21} = q^{-1} E_{31}
\end{array}
\ee
where $q = e^{3\vartheta/2}$, and $E_{ij} \ast E_{kl} = \delta_{jk}E_{il}$ 
for all other combinations. This product is equivalent to the usual product of 
matrices in the sense of (\ref{eqst}). A corresponding transformation map $S$ 
is determined by $S(E_{ii}) = E_{ii}$ and $S(E_{ij}) = q^{-1} E_{ij}$, 
$S(E_{ji}) = q E_{ji}$ for $i<j$. Obviously, the above construction can 
be applied to the universal enveloping algebra of every simple Lie 
algebra with a symplectic structure on its root space.

\section{Moyal deformation of gauge theory and Seiberg--Witten map}
\setcounter{equation}{0}
In this section $\A$ denotes the algebra of smooth functions on
$\mathbb{R}^{2n}$. Let $x^i$, $i=1, \ldots,2n$, be coordinate functions and 
$\pa_i$ the corresponding partial derivatives.
The Moyal product is defined as in the previous section with
\be
   P =  \theta^{ij}(\vartheta) \, \pa_i \otimes \pa_j
\ee
where $\theta^{ij}$ depends on a deformation parameter $\vartheta$, 
but not on the coordinates $x^i$. Hence 
$[x^i,x^j]_\ast = x^i \ast x^j - x^j \ast x^i = \theta^{ij}$.
Let $(\Omega(\A), \d)$ be the differential calculus over $(\A, \ast)$ such that 
\be
   [\d x^i,x^j]_\ast = \d x^i \ast x^j-x^j \ast \d x^i = 0  \; .
\ee
Using the Leibniz rule and $\d^2=0$, this implies
\be
   \d x^i \ast \d x^j + \d x^j \ast \d x^i = 0  \; .
\ee
In the limit $\vartheta \to 0$ we recover the ordinary differential calculus
on $\A$ with the usual product.
\vskip.1cm

Let $\psi$ transform according to 
\be
      \psi \mapsto \psi' = g \ast \psi      \label{psi-gt}
\ee
where $g$ is a map from $\mathbb{R}^{2n}$ into a representation of a Lie group $G$. 
The exterior covariant derivative of $\psi$ is
\be
    \D \psi = \d \psi + A \ast \psi
\ee
where $A = A_i \ast \d x^i$ is a matrix of 1-forms. It transforms in the same way 
as $\psi$, i.e., $D' \psi' = g \ast D \psi$, if the gauge potential transforms 
as follows,
\be
    \d g = g \ast A - A' \ast g  \; .        \label{ncgt}
\ee
One finds that
\be
    \D^2 \psi = F \ast \psi
\ee
with the field strength (or curvature)
\be
   F = \d A + A \ast A = {1 \over 2} \, F_{ij} \ast \d x^i \ast \d x^j
          \, , \qquad
   F_{ij} = \pa_i A_j - \pa_j A_i + [A_i,A_j]_\ast
          \label{cur}
\ee
which transforms as $F \mapsto F' = g \ast F \ast g^{-1}_\ast$ where $g^{-1}_\ast$ 
is the $\ast$-inverse of $g$.
\vskip.1cm

If a field $\varphi$ transforms as 
$\varphi \mapsto \varphi' = \varphi \ast g^{-1}_\ast$, 
its covariant derivative is $\D \varphi = \d \varphi - \varphi \ast A$ and we have 
$\D^2 \varphi = - \varphi \ast F$. Furthermore, if $B$ transforms as 
$B \mapsto B' = g \ast B \ast g^{-1}_\ast$, then 
$\D B = \d B + A \ast B - B \ast A$ and $\D^2 B = F \ast B - B \ast F$. Using (\ref{cur}) 
and the Leibniz rule for $\d$, we obtain the Bianchi identity $\D F = 0$.
\vskip.1cm

In the following we explore the $\vartheta$-dependence of the above gauge theoretical 
formulas. In particular, we are looking for a way to construct a noncommutative gauge 
transformation from an ordinary (commutative) one. First we note that
\be
   {\pa \over \pa \vartheta}(f \ast h) =
   {\pa f \over \pa \vartheta} \ast h + f \ast {\pa h \over \pa \vartheta}
   + \td f \wedge_\ast \td h  \; .    \label{aux-eq}
\ee
In particular, this shows that $\pa/\pa \vartheta$ is not a derivation of the 
$\ast$-product.
Here we introduced, as an auxiliary structure which greatly helps to simplify the 
following calculations, a differential calculus $(\tilde{\Omega}(\Omega(\A)),\td)$ 
over the algebra $\Omega(\A)$ where $\td$ is defined by\footnote{The ordinary
differentials $\d f$ and, moreover, all elements of $\Omega(\A)$ are 0-forms in $\tilde{\Omega}(\Omega(\A))$.}
\be
    \td f = \pa_i f \ast \td x^i \, , \qquad
    [ \td x^i , x^j]_\ast = 0 \, , \qquad
    \td (\d x^i) = 0  \; .
\ee
 Furthermore, we define an antisymmetric bilinear form on $\tilde{\Omega}^1$ 
by\footnote{This should not be confused with the wedge product in 
$\tilde{\Omega}(\Omega(\A))$, which is not needed in this work.}
\be
   \td x^i \wedge_\ast \td x^j = {1 \over 2} \, \epsilon^{ij}  \, , \qquad
   \epsilon^{ij} = {\pa \theta^{ij} \over \pa \vartheta}   \label{ab-form}
\ee
which means
\be
    \td f \wedge_\ast \td f' = {1 \over 2} \, \epsilon^{ij} \, \pa_i f \ast \pa_j f'
\ee
for $f,f' \in \A$. The map $\d$ extends to $\tilde{\Omega}(\Omega(\A))$
as a linear map if we require $\d \td = \td \d$.
\vskip.1cm

Differentiation of (\ref{psi-gt}) with respect to $\vartheta$ leads to
\be
   {\pa \psi' \over \pa \vartheta} = {\pa g \over \pa \vartheta} \ast \psi 
  + g \ast {\pa \psi \over \pa \vartheta} + \td g \wedge_\ast \td \psi \; .
\ee
This can be rewritten as follows,
\be
    {\pa \psi' \over \pa \vartheta} + \tA' \wedge_\ast \td \psi'
  = g \ast ( {\pa \psi \over \pa \vartheta} + \tA \wedge_\ast \td \psi ) 
  + ({\pa g \over \pa \vartheta} + \tA' \wedge_\ast \td g ) \ast \psi
           \label{psi_vartheta-gt1}
\ee
where we used
\be
  \td g = g \ast \tA - \tA' \ast g  \, , \qquad \tA = A_i \ast \td x^i
                     \label{tftg}
\ee
(which follows from (\ref{ncgt})), the Leibniz rule for $\td$, and (\ref{psi-gt}). 
Introducing a matrix field $\Gamma$ with the transformation law 
\be
   {\pa g \over \pa \vartheta} + \tA' \wedge_\ast \td g  
  = g \ast \Gamma - \Gamma' \ast g   \, ,    \label{Gamma'}
\ee
(\ref{psi_vartheta-gt1}) becomes
\be
    \nabla'_\vartheta \psi' = g \ast \nabla_\vartheta \psi
\ee
with the covariant derivative
\be
   \nabla_\vartheta \psi = {\pa \psi \over \pa \vartheta} 
            + \tA \wedge_\ast \td \psi + \Gamma \ast \psi  \; .
            \label{nabla-psi}
\ee
 For a field $\varphi$ with $\varphi' = \varphi \ast g^{-1}_\ast$, an analogous 
calculation leads to the covariant derivative
\be
   \nabla_\vartheta \varphi = {\pa \varphi \over \pa \vartheta} 
      - \td \varphi \wedge_\ast \tA - \varphi \ast \Lambda   \label{nabla_phi}
\ee
with a matrix field $\Lambda$ which transforms as follows,
\be
    {\pa g \over \pa \vartheta} - \td g \wedge_\ast \tA'
  = g \ast \Lambda - \Lambda' \ast g \; .  \label{Lambda'} 
\ee
Together with (\ref{tftg}) and (\ref{Gamma'}), the last equation leads to
\be
    g \ast ( \Gamma - \Lambda - \tA \wedge_\ast \tA ) 
	= ( \Gamma' - \Lambda' - \tA' \wedge_\ast \tA' ) \ast g  \; .
\ee
Since $Q = \Gamma - \Lambda - \tA \wedge_\ast \tA$ can be absorbed via a 
redefinition of $\Lambda$ in (\ref{Lambda'}), we are allowed to set $Q=0$ 
and get the following relation between $\Gamma$ and $\Lambda$,
\be
    \Lambda = \Gamma - \tA \wedge_\ast \tA  \; .    
\ee
Inserting this expression for $\Lambda$ in (\ref{nabla_phi}), we find
\be
  \nabla_\vartheta \varphi = {\pa \varphi \over \pa \vartheta} 
      - \tilde{\D} \varphi \wedge_\ast \tA - \varphi \ast \Gamma  \; .
\ee
A more symmetric form for the covariant derivatives of $\psi$ and $\varphi$ 
is achieved by setting
\be
    \Gamma = {1 \over 2} \, \tA \wedge_\ast \tA + \gamma \, , \qquad
	\Lambda = - {1 \over 2} \, \tA \wedge_\ast \tA + \gamma 
\ee
with a matrix field $\gamma$. Then
\be
  \nabla_\vartheta \psi &=& {\pa \psi \over \pa \vartheta} 
    + {1 \over 2} \, \tA \wedge_\ast (\td \psi + \tilde{\D} \psi ) + \gamma \ast \psi  \\
  \nabla_\vartheta \varphi &=& {\pa \varphi \over \pa \vartheta} 
    - {1 \over 2} \, ( \td \varphi + \tilde{\D} \varphi ) \wedge_\ast \tA 
	- \varphi \ast \gamma  \; .
\ee
 For a field $B$ with $B' = g \ast B \ast g^{-1}_\ast$, similar calculations lead to 
the covariant derivative
\be
      \nabla_\vartheta B 
  &=& {\pa B \over \pa \vartheta} + \tA \wedge_\ast \td B - \td B \wedge_\ast \tA
      - \tA \ast B \wedge_\ast \tA + \Gamma \ast B - B \ast \Lambda  \nonumber \\  
  &=& {\pa B \over \pa \vartheta} + \tA \wedge_\ast \td B - \tilde{\D} B \wedge_\ast \tA
      + \Gamma \ast B - B \ast \Gamma  \nonumber \\ 
  &=& {\pa B \over \pa \vartheta} + {1 \over 2} \, \tA \wedge_\ast ( \td B + \tilde{\D} B )
      - {1 \over 2} \, ( \td B + \tilde{\D} B ) \wedge_\ast \tA 
	  + \gamma \ast B - B \ast \gamma  \; .    \label{nabla-B}
\ee
 Furthermore, we have
\be
   \nabla_\vartheta (B \ast \psi) = (\nabla_\vartheta B) \ast \psi 
 + B \ast \nabla_\vartheta \psi + \tilde{\D} B \wedge_\ast \tilde{\D} \psi \, ,
              \label{def_deriv}
\ee
for example, which is a covariant version of (\ref{aux-eq}). The last term destroys 
the familiar ``derivation" property of covariant derivatives.
\vskip.1cm

Differentiation of (\ref{ncgt}) with respect to $\vartheta$ and use 
of (\ref{tftg}) and (\ref{Gamma'}) leads to 
\be
    \nabla'_\vartheta A' = g \ast (\nabla_\vartheta A) \ast g^{-1}_\ast
\ee
where
\be
       \nabla_\vartheta A 
   &=& {\pa A \over \pa \vartheta} + \tA \wedge_\ast \td A - \tilde{F} \wedge_\ast \tA 
        - \D \Gamma  \nonumber \\ 
   &=& {\pa A \over \pa \vartheta} + {1 \over 2} \, \tA \wedge_\ast 
       ( \td A + \tilde{F} ) - {1 \over 2} \, ( \td A + \tilde{F} ) \wedge_\ast \tA 
       - \D \gamma  
\ee 
with $\D \Gamma = \d \Gamma + A \ast \Gamma - \Gamma \ast A$, a corresponding definition
for $\D \gamma$, and 
\be
    \tilde{F} = \td A - \d \tA + \tA \ast A - A \ast \tA  \; .
\ee
\vskip.1cm

Moreover, we have
\be
     \nabla_\vartheta (\D B) - \D \nabla_\vartheta B 
   = \tilde{F} \wedge_\ast \tilde{\D} B - \tilde{\D} B \wedge_\ast \tilde{F}
     + (\nabla_\vartheta A) \ast B - B \ast \nabla_\vartheta A  
\ee
from which the corresponding formulas for $\nabla_\vartheta (\D \psi)$ and 
$\nabla_\vartheta (\D \varphi)$, for example, are evident. In particular,
\be
   \nabla_\vartheta (\D^2 \psi) = \D (\nabla_\vartheta \D \psi) 
  + \tilde{F} \wedge_\ast \tilde{\D} \D \psi + (\nabla_\vartheta A) \ast \D \psi \; .
\ee
Using $\nabla_\vartheta \D \psi = \D \nabla_\vartheta \psi 
+ \tilde{F} \wedge_\ast \tilde{\D} \psi + (\nabla_\vartheta A) \ast \psi$ and 
the Leibniz rule for $\D$, we obtain 
\be
   \nabla_\vartheta F = \D (\nabla_\vartheta A) + \tilde{F} \wedge_\ast \tilde{F} \; .
                           \label{nablaF}
\ee
\vskip.1cm

If we require $\nabla_\vartheta A=0$, which means
\be
   \pat{A} = - {1 \over 2} [\tA \wedge_\ast (\td A + \tilde{F})
             - (\td A + \tilde{F}) \wedge_\ast \tA] + \D \gamma  \; ,   
                      \label{swA} 
\ee
then we have also $\nabla'_\vartheta A'=0$ and thus
\be
   \pat{A}' = - {1 \over 2} [\tA' \wedge_\ast (\td A' + \tilde{F}')
           - (\td A' + \tilde{F}') \wedge_\ast \tA'] + \D' \gamma'  \; .   
                      \label{swA'} 
\ee
Together with (\ref{Gamma'}) which reads 
\be
   \pat{g} = {1 \over 2} (\td g \wedge_\ast \tA - \tA' \wedge_\ast \td g) 
             + g \ast \gamma - \gamma' \ast g  \, ,   \label{swg} 
\ee
this forms a system of first order differential equations 
which determines $g(\vartheta), A(\vartheta)$ and $A'(\vartheta)$ (here we 
suppress the dependence on the coordinates $x^i$, for simplicity) from 
$g(0), A(0)$ and $A'(0)$ and a choice of $\gamma$ and $\gamma'$.\footnote{We may 
discard the equation for $A'$ and eliminate $A'$ in (\ref{swg}) using (\ref{ncgt}).}
This means that, given a classical gauge transformation, the above 
equations determine a corresponding noncommutative gauge 
transformation. 
This is a {\em Seiberg-Witten map} \cite{Seib+Witt99}.
In particular, expanding $g$, $A$ and $A'$ in powers of $\vartheta$, the 
coefficients of the $(n+1)$th power are determined via (\ref{swA}), (\ref{swA'}) 
and (\ref{swg}) by the $n$th order coefficients and thus recursively by 
the 0th order.  
\vskip.1cm

Using $\nabla_\vartheta A=0$ and (\ref{nabla-B}) in (\ref{nablaF}) yields
\be
   \pat{F} = \tilde{F} \wedge_\ast \tilde{F}
             - {1 \over 2} \, [\tA \wedge_\ast (\td F + \tilde{\D} F)
             - (\td F + \tilde{\D} F) \wedge_\ast \tA] 
             - \gamma \ast F + F \ast \gamma  \; .
               \label{swF}
\ee
This first order differential equation determines the curvature of the 
noncommutative connection from that of the commutative connection $A$ 
at $\vartheta=0$. In particular, $F(0)=0$ implies $F(\vartheta)=0$ 
for all $\vartheta$.
\vskip.1cm

Expressed in components, the equations (\ref{swg}), (\ref{swA}) and (\ref{swF}) 
with $\gamma=0=\gamma'$ take the form
\be
  {\pa g \over \pa \vartheta} &=& {1 \over 4} \, \epsilon^{ij} \, ( \pa_i g \ast A_j 
                                  + A_j' \ast \pa_i g )   \\
  {\pa A_i \over \pa \vartheta} &=& - {1 \over 4} \, \epsilon^{kl} \, 
                   [A_k , \pa_l A_i + F_{li}]_{\ast,+}  \\
  {\pa F_{ij} \over \pa \vartheta} &=& {1 \over 4} \,\epsilon^{kl} \, 
     (2 [F_{ik} , F_{jl}]_{\ast,+} - [A_k,D_l F_{ij} + \pa_l F_{ij}]_{\ast,+})
\ee
where $[f , h]_{\ast,+} = f \ast h + h \ast f$. From these equations one recovers 
equations (3.8) in \cite{Seib+Witt99} for an infinitesimal gauge transformation. 
\vskip.1cm

We can extend the Seiberg-Witten map to matter fields like $\psi, \varphi, B$ by 
setting their covariant $\vartheta$-derivatives to zero. For example, 
$\nabla_\vartheta \psi =0$ with $\gamma=0$ leads to
\be
   {\pa \psi \over \pa \vartheta} 
 = - {1 \over 4} \, \epsilon^{ij} \, A_i \ast ( \pa_j \psi + \nabla_j \psi ) \; .
\ee

\section{Parallel transport in deformation space and curvature}
\setcounter{equation}{0}
In general, $\theta^{ij}$ may depend on several deformation parameters $\vartheta_i$. 
The deformation elaborated in section 3 can then be performed along any curve in 
the deformation space $\Theta$ on which $\theta^{ij}$ are functions. We learned 
that a Seiberg-Witten map has the geometric interpretation of a parallel 
transport along a curve in $\Theta$. In general, such a parallel transport is 
path-dependent due to the presence of a curvature associated with the covariant 
derivatives (see also \cite{Asak+Kish99}). 
\vskip.1cm

Instead of $\td$, $\tA$, $\Gamma$ (and other quantities) which refer to a deformation 
parameter $\vartheta$, we write $\d_1$, $A_1$, $\Gamma_1$ and $\d_2$, $A_2$, $\Gamma_2$, 
refering to deformation parameters $\vartheta_1$ and $\vartheta_2$, respectively. 
Correspondingly, there are two different antisymmetric bilinear forms replacing 
(\ref{ab-form}) with $\epsilon_1^{ij} = \pa \theta^{ij} / \pa \vartheta_1$ and 
$\epsilon_2^{ij} = \pa \theta^{ij} / \pa \vartheta_2$. In particular, 
\be
  \nabla_{\vartheta_1} \psi &=& {\pa \psi \over \pa \vartheta_1 } 
   + A_1 \wedge_\ast \d_1 \psi + \Gamma_1 \ast \psi \, , \\
  \nabla_{\vartheta_2} \psi &=& {\pa \psi \over \pa \vartheta_2 } 
   + A_2 \wedge_\ast \d_2 \psi + \Gamma_2 \ast \psi
\ee
replace (\ref{nabla-psi}). Now 
\be
     [ \nabla_{\vartheta_1} , \nabla_{\vartheta_2} ] \psi 
 &=& F_{12} \wedge_\ast \d_1 \d_2 \psi + (F_{12} \wedge_\ast A_2 
     - \nabla_{\vartheta_2} A_1) \wedge_\ast \d_1 \psi  
	            \nonumber \\
 & & + (F_{12} \wedge_\ast A_1 + \nabla_{\vartheta_1} A_2) \wedge_\ast \d_2 \psi
	            \nonumber \\
 & & + ( K_{12} + A_1 \wedge_\ast \d_1 \Gamma_2 - A_2 \wedge_\ast \d_2 \Gamma_1 ) \ast \psi
\ee
where
\be
    F_{12} &=& \d_1 A_2 - \d_2 A_1 + A_1 \ast A_2 - A_2 \ast A_1 \, , \\
    K_{12} &=& {\pa \over \pa \vartheta_1} \Gamma_2 - {\pa \over \pa \vartheta_2} \Gamma_1 
	       + \Gamma_1 \ast \Gamma_2 - \Gamma_2 \ast \Gamma_1 \; .
\ee
After several manipulations, one arrives at the following generalized Ricci identity,
\be
    [ \nabla_{\vartheta_1} , \nabla_{\vartheta_2} ] \psi &=& {\cal F}_{12} \ast \psi
	    + {1 \over 2} \, F_{12} \wedge_\ast (\D_1 \D_2 + \D_2 \D_1) \psi \nonumber \\
	& & + (\nabla_{\vartheta_1} A_2) \wedge_\ast \D_2 \psi 
	    - (\nabla_{\vartheta_2} A_1) \wedge_\ast \D_1 \psi   \label{nabla12-psi}
\ee
which is evidently covariant since the {\em generalized curvature} 
\be
        {\cal F}_{12} 
    &=& K_{12} + A_1 \wedge_\ast \d_1 \Gamma_2 - A_2 \wedge_\ast \d_2 \Gamma_1          
        + (\nabla_{\vartheta_2} A_1) \wedge_\ast A_1 
		- (\nabla_{\vartheta_1} A_2) \wedge_\ast A_2        \nonumber \\
    & & - {1 \over 2} \, F_{12} \wedge_\ast ( \d_1 A_2 + \d_2 A_1 + A_1 \ast A_2 + A_2 \ast A_1 ) 
		                       \nonumber \\
    &=& K_{12}
        + {1 \over 2} \, {\pa \theta^{ij} \over \pa \vartheta_1} \, A_i \ast \pa_j \Gamma_2  
	    - {1 \over 2} \, {\pa \theta^{ij} \over \pa \vartheta_2} \, A_i \ast \pa_j \Gamma_1
		     \nonumber \\
    & &	+ {1 \over 2} \, {\pa \theta^{ij} \over \pa \vartheta_1} \, 
          (\nabla_{\vartheta_2} A_i) \ast A_j    
        - {1 \over 2} \, {\pa \theta^{ij} \over \pa \vartheta_2} \, 
          (\nabla_{\vartheta_1} A_i) \ast A_j 		     \nonumber \\
    & & - {1 \over 8} \, {\pa \theta^{ij} \over \pa \vartheta_1} \, 
          {\pa \theta^{kl} \over \pa \vartheta_2} 
          \, F_{ik} \ast ( \pa_j A_l + \pa_l A_j + A_j \ast A_l + A_l \ast A_j )						   
\ee
transforms as follows,
\be
    {\cal F}'_{12} = g \ast {\cal F}_{12} \ast g^{-1}_\ast  \; .
\ee
This result is obtained by a lengthy calculation starting with
\be
    \left( {\pa \over \pa \vartheta_1} {\pa \over \pa \vartheta_2} 
  - {\pa \over \pa \vartheta_2} {\pa \over \pa \vartheta_1} \right) g = 0 \, ,
\ee
using (\ref{Gamma'}) in the form
\be
   {\pa g \over \pa \vartheta} + A'_1 \ast g \wedge_\ast A_1
  = g \ast \Gamma_1 - \Lambda'_1 \ast g 
\ee
and correspondingly with the index 1 replaced by 2, and noting that the generalized 
curvature also has the following expression,
\be
        {\cal F}_{12} 
    &=& {\pa \over \pa \vartheta_1} \Lambda_2 - {\pa \over \pa \vartheta_2} \Lambda_1 
	       + \Lambda_1 \ast \Lambda_2 - \Lambda_2 \ast \Lambda_1
		   - \d_1 \Lambda_2 \wedge_\ast A_1 + \d_2 \Lambda_1 \wedge_\ast A_2 \nonumber \\         
    & & - A_1 \wedge_\ast \nabla_{\vartheta_2} A_1
		+ A_2 \wedge_\ast \nabla_{\vartheta_1} A_2       \nonumber \\
    & & + {1 \over 2} \, ( \d_1 A_2 + \d_2 A_1 - A_1 \ast A_2 - A_2 \ast A_1 ) 
	    \wedge_\ast F_{12} \; .						   
\ee
Besides the generalized curvature, there are additional terms on the rhs of the 
Ricci identity (\ref{nabla12-psi}). Their origin lies in the deviation of the 
covariant derivative $\nabla_\vartheta$ from a ``derivation" (cf (\ref{def_deriv})). 
\vskip.1cm

The formula which replaces (\ref{nabla12-psi}) for the field $\varphi$ is
\be
      [ \nabla_{\vartheta_1} , \nabla_{\vartheta_2} ] \varphi 
  &=& - \varphi \ast {\cal F}_{12} - {1 \over 2} \, [(\D_1 \D_2 + \D_2 \D_1) \varphi] \wedge_\ast F_{12}
            \nonumber \\
	& & + \D_1 \varphi \wedge_\ast \nabla_{\vartheta_2} A_1 
	    - \D_2 \varphi \wedge_\ast \nabla_{\vartheta_1} A_2  \; .  
\ee
\vskip.1cm

The path-dependence of Seiberg-Witten maps leads to the following idea. 
Consider a closed path through $\vartheta=0$ in the deformation parameter space. 
We could imagine that, as a consequence of the nonvanishing Ricci identity, 
parallel transport from $\vartheta=0$ along the path back to $\vartheta=0$ 
maps a solution of some commutative model to another solution. This may lead 
to a solution generating method.

\section{Moyal deformations of integrable models and Seiberg-Witten map}
\setcounter{equation}{0}
Let $(\Omega,\d,\dl)$ be a bi-differential calculus \cite{DMH00a} over $(\A,\ast)$ 
such that the bicomplex conditions $\d^2 = \delta^2 = \d \delta + \delta \d =0$ 
are identically satisfied. Replacing $\d$ with $\D$ defined by 
\be
  \D \psi = g^{-1}_\ast \ast \d(g \ast \psi) = \d \psi + A \ast \psi \, , \qquad
        A = g^{-1}_\ast \ast \d g    \label{dress}
\ee
for some $\ast$-invertible matrix-valued function $g$, the new bicomplex 
conditions are equivalent to
\be
     \dl A =0  \; .    \label{delta=0}
\ee
Note that, as a consequence of the definition of $A$, the curvature of $A$ 
vanishes, i.e.,
\be
   F = \d A + A \ast A = 0  \; .
\ee
 For $\vartheta=0$, such a bicomplex can be associated with many integrable models, 
including the self-dual Yang-Mills equations, in such a way that (\ref{delta=0}) 
is equivalent to the integrable model equation \cite{DMH00a}. 
\vskip.1cm

Is it possible to obtain solutions of the noncommutative version from those of 
a classical integrable model via the Seiberg-Witten map? Indeed, we already know 
that the Seiberg-Witten map preserves the zero curvature condition.\footnote{Most 
integrable models admit a zero curvature formulation with a parameter ($\lambda$) 
dependent connection. Since the Seiberg-Witten map is quadratic in the connection, 
it does not, in general, respect the concrete $\lambda$-dependence of a flat 
connection. This results in constraints and thus obstructions to construct 
solutions of the deformed model from the commutative one.}  
It remains to investigate with the help of the formalism of section 3 whether 
this map also preserves (\ref{delta=0}). $\td$ extends to $\Omega$ with the 
additional rules $\delta \td = \td \delta$ and $\td \delta x^i =0$. Applying 
$\dl$ to (\ref{swA}), using $F=0$ and $\delta \td = \td \delta$, we get
\be
     {\pa \over \pa \vartheta} \dl A
  =  \dl {\pa A \over \pa \vartheta} 
  =  - {1 \over 2} [\dl \tA \wedge_\ast \td A + \tA \wedge_\ast \td \dl A
     - \td \dl A \wedge_\ast \tA + \td A \wedge_\ast \dl \tA]  \; .
\ee
If the additional condition
\be
   \dl \tA \wedge_\ast \td A  + \td A \wedge_\ast \dl \tA =0
                \label{con}
\ee
holds, then the last equation indeed implies $\dl A(\vartheta) = 0$ if 
$\dl A(0) = 0$. As a consequence, each solution of the classical integrable 
model generates a solution of the noncommutative version, if the latter 
solution satisfies (\ref{con}).
\vskip.1cm

As an example, let us start with the trivial bi-differential calculus 
determined by
\be
    \d \psi = \psi_t \ast \d t + \psi_x \ast \d x \, , \qquad
   \dl \psi = \psi_x \ast \d t + \psi_t \ast \d x
\ee
with coordinates $t$ and $x$, and ``dress" the first operator according to 
(\ref{dress}) so that 
\be
  \D \psi = (\psi_t + U \ast \psi) \, \d t + (\psi_x + V \ast \psi) \, \d x
\ee
where
\be
   U = g^{-1}_\ast \ast g_t \, , \qquad
   V = g^{-1}_\ast \ast g_x  \; .
\ee
The bicomplex conditions are satisfied if and only if
\be
   (g^{-1}_\ast \ast g_t)_t - (g^{-1}_\ast \ast g_x)_x = 0
                  \label{pcm-eq}
\ee
which is the noncommutative version of the principal chiral
field equation (see also \cite{Taka00}). The condition (\ref{con}) becomes
\be
   \dl \tA \wedge_\ast \td A  + \td A \wedge_\ast \dl \tA 
 = {1 \over 2} \, [ U_x - V_t , V_x - U_t ]_{\ast,+} \ast \d t \ast \d x
\ee 
which vanishes as a consequence of the field equation (\ref{pcm-eq}). 
Hence, every solution of the classical principal chiral model 
generates a solution of the noncommutative model. In practice, this 
allows at least the recursive calculation of the coefficients of a 
power series expansion of the field $g(\vartheta)$ in the deformation 
parameter $\vartheta$ from a given classical solution $g(0)$. Convergence 
of the resulting formal power series has still to be investigated.
\vskip.1cm

Using (\ref{ncgt}) and $t \ast x - x \ast t = \vartheta$, (\ref{swg}) leads to
\be
   {\pa g \over \pa \vartheta}
 = {1\over 4} (g_x \ast \pa_t g^{-1}_{\ast} - g_t \ast \pa_x g^{-1}_{\ast}) \ast g
 = -{1 \over 2} (\td g \wedge_\ast \td g^{-1}_\ast) \ast g
                                 \label{pc}
\ee
with $\td f = f_t \, \td t + f_x \, \td x$.
A special class of classical chiral models is defined by $g_0 = I-2 \Pi_0$ 
where $\Pi_0$ is a projection, i.e., $\Pi_0^2 = \Pi_0$.
Let us try to find a deformation of this particular class. If we assume that
$g = I-2 \Pi$ with $\Pi \ast \Pi = \Pi$, then we have $g^{-1}_\ast = I-2 \Pi$. 
Substituting this in (\ref{pc}), we find
\be
  {\pa \Pi \over \pa \vartheta} = (\td \Pi \wedge_\ast \td \Pi) \ast (I-2 \Pi) \; .
\ee
With the help of the equation derived from $\Pi \ast \Pi = \Pi$ by differentiation 
with respect to $\vartheta$, and using the Leibniz rule for $\td$, we obtain 
\be
    \td \Pi \wedge_\ast \td \Pi = 0
\ee
which is an additional condition for $\Pi$ if $\vartheta \neq 0$. 
This shows that the Seiberg-Witten map is not necessarily consistent with 
reductions of the principal chiral model.

\section{Final remarks}
\setcounter{equation}{0}
Let ncEQS stand for a (noncommutative) deformation of a system EQS of field 
equations. If we can find a system of first order differential equations 
in the deformation parameter $\vartheta$, as a consequence of which 
$\pa (ncEQS)/\pa \vartheta =0$, then (under certain technical conditions) 
solutions of ncEQS are obtained from solutions of EQS. The Seiberg-Witten 
map for gauge fields provides us with an example. It allows us, in particular, 
to construct solutions of the noncommutative zero curvature condition from 
solutions of the classical zero curvature condition. Moreover, we have shown 
that this map also works in case of the two-dimensional principal chiral model 
and its noncommutative version. Another example which fits into the above 
scheme is the noncommutative KdV equation treated in \cite{DMH00c}.

\end{document}